\pdfoutput=1
\documentclass[11pt,twoside,a4paper,cmspaper,final,collab]{cms-tdr}
\def\svnVersion{398daa2}\def\svnDate{2024/02/12}\def\cmsCernNoTag{CERN-EP-2022-219}\def\cmsCernDate{\today}\def\cmsMessage{Published in Physics Letters B as \href{http://dx.doi.org/10.1016/j.physletb.2023.138389}{\doi{10.1016/j.physletb.2023.138389}.}}
\begin{document}\cmsNoteHeader{HIN-21-003}

\newcommand {\minv}{\ensuremath{m_{\text{inv}}}\xspace}
\newcommand {\vn}{\ensuremath{v_{\mathrm{n}}}\xspace}
\newcommand{\vtwo}{\ensuremath{v_{2}}\xspace}
\newcommand {\vthree}{\ensuremath{v_{3}}\xspace}
\newcommand{\Dtopipi}{\ensuremath{\PDz\to\PGpp\PGpm}\xspace}
\newcommand{\DtoKK}{\ensuremath{\PDz\to\PKp\PKm}\xspace}
\newcommand{\PbPb}{\ensuremath{\text{PbPb}}\xspace}
\newcommand{\DCA}{\ensuremath{\text{DCA}}\xspace}
\newcommand{\QnDz}{\ensuremath{Q_{n}^{\PDz}}\xspace}
\newcommand{\QnA}{\ensuremath{Q_{n \mathrm{A}}}\xspace}
\newcommand{\QnB}{\ensuremath{Q_{n \mathrm{B}}}\xspace}
\newcommand{\QnC}{\ensuremath{Q_{n \mathrm{C}}}\xspace}
\newcommand{\wj}{\ensuremath{w_{j}}\xspace}
\newlength\cmsTabSkip\setlength{\cmsTabSkip}{1ex}

\cmsNoteHeader{HIN-21-003}
\title{Measurements of azimuthal anisotropy of nonprompt \texorpdfstring{\PDz}{D0} mesons in \texorpdfstring{\PbPb}{PbPb} collisions at \texorpdfstring{$\sqrtsNN=5.02\TeV$}{sqrt(sNN) = 5.02 TeV}}

\date{\today}

\abstract{
Measurements of the elliptic (\vtwo) and triangular (\vthree) azimuthal anisotropy coefficients are presented for \PDz mesons produced in \PQb hadron decays (nonprompt \PDz mesons) in lead-lead collisions at $\sqrtsNN=5.02\TeV$. The results are compared with previously published charm meson anisotropies measured using prompt \PDz mesons. The data were collected with the CMS detector in 2018 with an integrated luminosity of 0.58\nbinv. Azimuthal anisotropy is sensitive to the interactions of quarks with the hot and dense medium created in heavy ion collisions. Comparing results for prompt and nonprompt \PDz mesons can assist in understanding the mass dependence of these interactions.
The nonprompt results show lower magnitudes of \vtwo and \vthree and weaker dependences on the meson transverse momentum and collision centrality than those found for prompt \PDz mesons. 
The results are in agreement with theoretical predictions that include a mass dependence in the interactions of quarks with the medium.
}

\hypersetup{%
pdfauthor={CMS Collaboration},%
pdftitle={Azimuthal anisotropy of nonprompt D0 mesons in PbPb collisions at sqrt(sNN) = 5.02 TeV},%
pdfsubject={CMS},%
pdfkeywords={CMS, heavy ions, QPG, flow, heavy flavour}}

\maketitle 

\section{Introduction}
\label{sec:intro}

In ultra-relativistic heavy ion collisions,
 cold nuclear matter transforms into a state of strongly coupled matter where quarks and gluons are deconfined from
hadrons, called the quark-gluon plasma (QGP)~\cite{BRAHAMS:2005,PHOBOS:2005,STAR:2005,PHENIX:2005}. 
One of the main features of the QGP is the collective motion of its
constituents. 
A signature of this collective motion is the presence of
a momentum anisotropy 
of the particles that
emerge from the collision. 
In the plane perpendicular to the beam axis (transverse plane), this anisotropy results from the conversion to momentum space of the initial transverse spatial anisotropy of the colliding system,
a phenomenon known as collective anisotropic flow.
To quantify this effect, azimuthal (Fourier) coefficients of order $n$, $\vn=\langle\cos [n (\phi - \Psi_{n})] \rangle$ can be used, where $\phi$ is the single-particle azimuthal angle and 
$\Psi_n$ is the first angle where the $n$th harmonic component has its maximum outgoing particle density~\cite{Voloshin:1996zpc,Busza:2018bp}.
The measured \vn values for low-momentum light hadrons can be 
described by relativistic hydrodynamics~\cite{Ollitrault:1992bk,Heinz:2013th,Gale:2013da}.
The second Fourier coefficient, \vtwo (referred to as elliptic flow), 
arises from the combined effects of the elongated shape and the event-by-event fluctuations of the overlap region of the colliding nuclei~\cite{Alver:2010prc}.
The third coefficient, \vthree (referred to as triangular flow), is predominantly due to the fluctuations.

Because of their large mass, bottom (\PQb) and charm (\PQc) quarks  are produced at the earliest stage of the collision~\cite{Prino:2016hf},
but their azimuthal distributions can be affected by their interaction with the medium as they travel through the QGP~\cite{Beraudo:2018enl}.
At low transverse momentum (\pt), it is
expected that charm quarks thermalize and, therefore, they should follow the motion of the gluons and light flavor quarks~\cite{vanHes:2006therm}. 
In the high \pt region,
the azimuthal asymmetry could be caused by a path length dependence of the parton energy loss~\cite{Gyulassy2001hptv2,Shuryak2002hptv2}. 
Hence, studying the flow coefficients of heavy quarks in heavy ion collisions can provide important inputs for understanding the properties of the QGP.
Significant \vtwo and \vthree coefficients of charm hadrons in lead-lead (\PbPb) collisions have already been measured at the CERN LHC, proving that charm quarks exhibit collective flow~\cite{Acharya:2018dvn,Sirunyan:2017plt,Sirunyan:2020dvn,Aad:2020bvn,Acharya:2020jpsivn,Acharya:2021Dvn,Sirunyan:2022dv24}. 
An early measurement by the CMS Collaboration of the elliptic flow of \cPJgy from \PQb hadron decays was consistent with zero, 
albeit with large uncertainties~\cite{Khachatryan:2017jps}.
Recently, the ALICE Collaboration measured the \vtwo of electrons from \PQb hadron decays~\cite{Acharya:2021bvn}, and the ATLAS Collaboration measured both \vtwo and \vthree of muons from \PQb hadron decays~\cite{Aad:2020bvn}.
These results indicate that \PQb hadron elliptic flow is nonzero, and also show a clear mass ordering,
with smaller measured values of Fourier coefficients than those found for particles containing lighter flavor quarks.
Compared to these \cPJgy and lepton measurements, the study of the $\PQb\to\PDz$ channel is a promising opportunity due to the large branching fraction of this decay, approximately 70\%~\cite{Workman:2022pdg}. 
This enables lower \pt coverage, thereby increasing sensitivity to the low-\pt flow of \PQb hadrons.
In addition, because of its large mass, the \PDz momentum is more closely correlated with the \PQb quark momentum than is the case for leptons.

In this Letter, the first \vtwo and \vthree measurements of \PDz mesons 
from \PQb hadron decay (nonprompt \PDz mesons) in large collision systems are reported, using \PbPb collisions at a center-of-mass energy per nucleon pair of $\sqrtsNN=5.02\TeV$. 
Results for \PDz mesons in the rapidity range $\abs{y}<1$ are shown as a function of their \pt, spanning 1--30\GeVc, and in three classes of \PbPb centrality: 0--10\%, 10--30\%, and 30--50\%. 
These centrality percentages represent the fraction of the total inelastic hadronic cross section, with 0\% corresponding to full overlap of the two colliding nuclei.
The broad range in \pt enables a comprehensive study of different flow generation mechanisms.  
The results are compared with azimuthal anisotropy measurements of \PDz mesons that do not come from \PQb hadron decays (prompt \PDz mesons)~\cite{Sirunyan:2020dvn}, 
as well as with theoretical predictions.
Tabulated results are provided in the HEPData record for this analysis~\cite{hepdata}.

\section{Experimental setup and data sample}
\label{sec:exp}

The CMS apparatus~\cite{Chatrchyan:2008zzk} is a multipurpose, nearly hermetic detector, designed to trigger on~\cite{CMS:2020cmk,CMS:2016ngn} and identify electrons, muons, photons, and (charged and neutral) hadrons~\cite{CMS:2020uim,CMS:2018rym,CMS:2014pgm}. A global algorithm~\cite{CMS:2017yfk} aims to reconstruct all individual particles in an event, combining information provided by the all-silicon inner tracker and by the crystal electromagnetic and brass-scintillator hadron calorimeters, operating inside a 3.8\unit{T} superconducting solenoid, with data from the gas-ionization muon detectors embedded in the flux-return yoke outside the solenoid. 
The forward hadron (HF) calorimeter uses steel as an absorber and quartz fibers as the sensitive material. 
The two halves of the HF are located 11.2\unit{m} from the interaction region, one on each end, and together they provide coverage in the pseudorapidity range $3.0<\abs{\eta}<5.2$. 
The HF calorimeters are subdivided into ``towers" with $\Delta \eta{\times} \Delta \phi = 0.175{\times} 0.175$, and energy deposited in a tower is treated as a detected hadron in this analysis.
They serve as luminosity monitors and 
the total energy deposited in both HF calorimeters is used for centrality determination~\cite{CMS:2011iwn}.
Events are filtered using a two-tiered trigger system. 
The first level, composed of custom hardware processors, uses information from the calorimeters and muon detectors~\cite{CMS:2020cmk}. 
The second level, known as the high-level trigger~\cite{CMS:2016ngn}, consists of a farm of processors running a version of the full event reconstruction software.

The data analyzed in this Letter consist of minimum bias (MB) PbPb events corresponding to an integrated luminosity of 0.58\nbinv~\cite{CMS:2021xjt,CMS-PAS-LUM-18-001}. 
The MB selection requires signals above readout thresholds in the range of ${\sim}6\text{--}12\GeV$ on both sides of the HF calorimeters~\cite{CMS:2016ngn}. 
The events are further filtered to remove background events (beam-gas interactions and nonhadronic collisions) by applying the procedure described in Ref.~\cite{Khachatryan:2016odn}. 
The final results include only events with at least one vertex associated with two or more tracks (primary vertex) within 15\cm from the nominal interaction point along the beam direction, 
and at least two towers with energy larger or equal to 4\GeV in each of the HF detectors. 
Finally, in order to suppress the contamination from events with multiple ion collisions, 
an event is rejected if the shapes of the clusters in the pixel part of the inner tracker are not compatible with those expected in single PbPb collisions. 
The total event selection efficiency is $(97\pm2)\%$.

Simulated events from Monte Carlo (MC) generators are used to study the kinematics of \PDz mesons in \PbPb collisions. 
The \PDz mesons are generated with \PYTHIA 8.212~\cite{Sjostrand:2014zea}, tune CP5~\cite{Sirunyan:2019dfx}, while their decays are modeled with \EVTGEN 1.3~\cite{Lange:2001uf}. 
Both prompt and nonprompt \PDz samples are produced. 
The decay products of the \PDz mesons are then embedded into MB events generated using \HYDJET 1.9~\cite{Lokhtin:2005px}, tuned to reproduce the charged particle multiplicity and spectra. 
The response of the CMS detector to these combined events is simulated with 
\GEANTfour~\cite{Agostinelli:2002hh}. 

\section{Analysis procedure}
\label{sec:ana}
 
Inclusive (both prompt and nonprompt) \PDz and \PADz meson candidates are reconstructed via their decay channels:
$\PDz\to\PGpp+\PKm$ and $\PADz\to\PGpm+\PKp$. 
Both \PDz and \PADz are referred to as \PDz in this Letter. 
All tracks used in the analysis have $\pt>1\GeVc$ and $\abs{\eta}<1.2$.
To ensure the best track quality, only tracks that satisfy high purity~\cite{CMS:2014pgm} criteria are considered. In addition, 
the relative uncertainty in the track \pt measurement is required to be less than 10\%.
The number of hits along the track trajectory in the tracker must be greater or equal to 11. 
The $\chi^{2}$ of the track fit divided by the total number of degrees of freedom of the fit, 
and also divided by the total number of tracker layers having a hit along the track path, must be less than 0.18.
Candidates are formed by combining pairs of tracks from oppositely-charged particles and requiring
an invariant mass ({\minv}) within a ${\pm}200\MeVcc$ window of the world-average 
\PDz meson mass of 1865\MeVcc~\cite{Workman:2022pdg}.
Since no particle identification is available, both possible particle assignments (pion and kaon)  are considered for each pair of selected candidate decay tracks.

Further selection is done by applying a boosted decision tree (BDT)
algorithm implemented using the \textsc{tmva} package~\cite{Hocker:2007ht}.
For the BDT training, the signal candidates are taken from the simulated events 
in which reconstructed \PDz mesons are required
to match the generated nonprompt \PDz particles. 
The background training sample consists of two components combined together: combinatorial background and prompt \PDz production.
The ratio of these two components in the background training sample is the same as found in the data.
The combinatorial background sample is composed of \PDz candidates reconstructed from data whose mass is three to six standard deviations away from the nominal \PDz meson mass (${\sim}1.795\text{--}1.830$ and ${\sim}1.900\text{--}1.935\GeVcc$).
The prompt \PDz meson component of the background sample consists  of \PDz candidates that correspond to simulated prompt \PDz mesons in the MC events.
In this way, the training is optimized to favor the \PDz mesons produced in \PQb hadron decays. 
Training is performed separately for each \pt range and centrality class. 
The variables related to \PDz mesons used to discriminate the signal from the background are: 
$\chi^2$ probability for the \PDz vertex fit, the distance between the \PDz and primary vertices and its significance, 
as well as the angle between 
the momentum of the \PDz meson candidate and the line connecting the primary and 
\PDz vertices (pointing angle). 
In addition, related to the decay products of the \PDz meson candidate, the variables used are:
the significance of the distances of closest approach (\DCA) to the primary vertex (both along and perpendicular to the beam direction),
and the number of hits 
assigned to a particular track.
These variables are chosen by analyzing their importance in the decision tree and the correlation matrix among all variables. 
The BDT cut is optimized to provide the maximal nonprompt \PDz meson signal significance for each analysis bin. 
Each \PDz candidate is weighted by a correction factor,
defined as the inverse of the product of the acceptance and selection efficiencies as functions of \pt and centrality. 
This factor is calculated from the MC sample using the ratio of the number of reconstructed \PDz mesons (that satisfy selection criteria) to the total
number of generated nonprompt \PDz mesons in a certain kinematic range.
Potential discrepancies between data and simulated \pt spectra could lead to an inaccuracy in the correction factor. 
Therefore, before performing this correction, the \pt spectrum of \PDz mesons in the simulation is weighted to match that in the data.
The weighting factor is calculated in each centrality range from the ratio of the uncorrected \pt distribution of nonprompt \PDz mesons, 
extracted from the invariant mass and \DCA template fits to data, and the model distribution, obtained with the same reconstruction and selection criteria applied to data.

{\tolerance=800 The measurement of inclusive \PDz meson \vn coefficients employs the scalar-product (SP) method~\cite{STAR:sp} used in previous CMS publications for prompt \PDz mesons~\cite{Sirunyan:2017plt,Sirunyan:2020dvn}.\par} 
In this method, the \vn coefficients of all \PDz candidates ($\vn^{\text{sig+bkg}}$) are determined using
\begin{linenomath*}
\begin{equation}
\label{eq:scalar_product_vn}
\centering
\vn \{ \mathrm{SP} \} \equiv \frac{ \langle \QnDz\QnA^{*} \rangle }{\sqrt{\smash[b]{\frac{ \langle \QnA\QnB^{*} \rangle \langle \QnA\QnC^{*} \rangle }{ \langle \QnB\QnC^{*} \rangle }}}},
\end{equation}
\end{linenomath*}
where the $Q$-vectors are defined as $Q_{n} \equiv \sum_{j=1} \wj \re^{in\phi_{j}}$,
the asterisk symbol (*) means complex conjugation, and the subscripts A, B, and C refer to different subevents.
In each event, the sum for \QnA and \QnB is over all hadrons detected in HF which are above a threshold energy,
while the sum for \QnC includes all reconstructed tracks above a \pt threshold.
The \QnDz signifies the flow vector of a \PDz meson candidate within a particular kinematic range.
The weight \wj is a dimensionless quantity and corresponds to the energy deposited in the HF tower in \GeV for \QnA and \QnB, or to the track \pt in \GeVc, at azimuthal angle $\phi_{j}$ for \QnC,
or $\wj=1$ in the case of \PDz meson candidates.
The subevent A (B) uses the negative (positive) side of the HF when
the \PDz meson candidate is at positive pseudorapidity, and vice versa.
This choice of subevents avoids autocorrelations and results in an $\eta$ gap
of at least three units between the \PDz meson daughters and particles from the underlying subevent, thereby suppressing short-range correlations.
Flattening and recentering procedures are applied to the $Q$-vectors related to HF and the tracker, for removing detector acceptance effects~\cite{Poskanzer:1998yz,Alt:2003ab}.
The averages
$\langle \QnA\QnB^{*} \rangle$, $\langle \QnA\QnC^{*} \rangle$, and
$\langle \QnB\QnC^{*} \rangle$ are found by considering all selected events, while the average
$\langle \QnDz\QnA^{*} \rangle$ includes all \PDz meson candidates in all selected events.

To extract the inclusive \PDz meson flow harmonics, the individual \PDz candidates are divided into bins of their flow vector SPs (\ie, Eq.~(\ref{eq:scalar_product_vn}) but not averaged over all \PDz candidates). 
A separate mass spectrum is generated in each of these SP bins and an invariant mass fit is performed to get the SP distribution of the \PDz signal. 
The inclusive \PDz \vn values can be obtained by:
\begin{linenomath*}
\begin{equation}
\label{eq:2dfit_vn}
\centering
\vn^{\PDz} =\frac{ \sum{\vn^i Y_i}}{\sum Y_i},
\end{equation}
\end{linenomath*}
where $\vn^i$ is the center of the $i$th SP bin, and the $Y_i$ is the \PDz yield in the same bin.
\begin{figure*}[htb]
\centering
\includegraphics[width=0.45\textwidth]{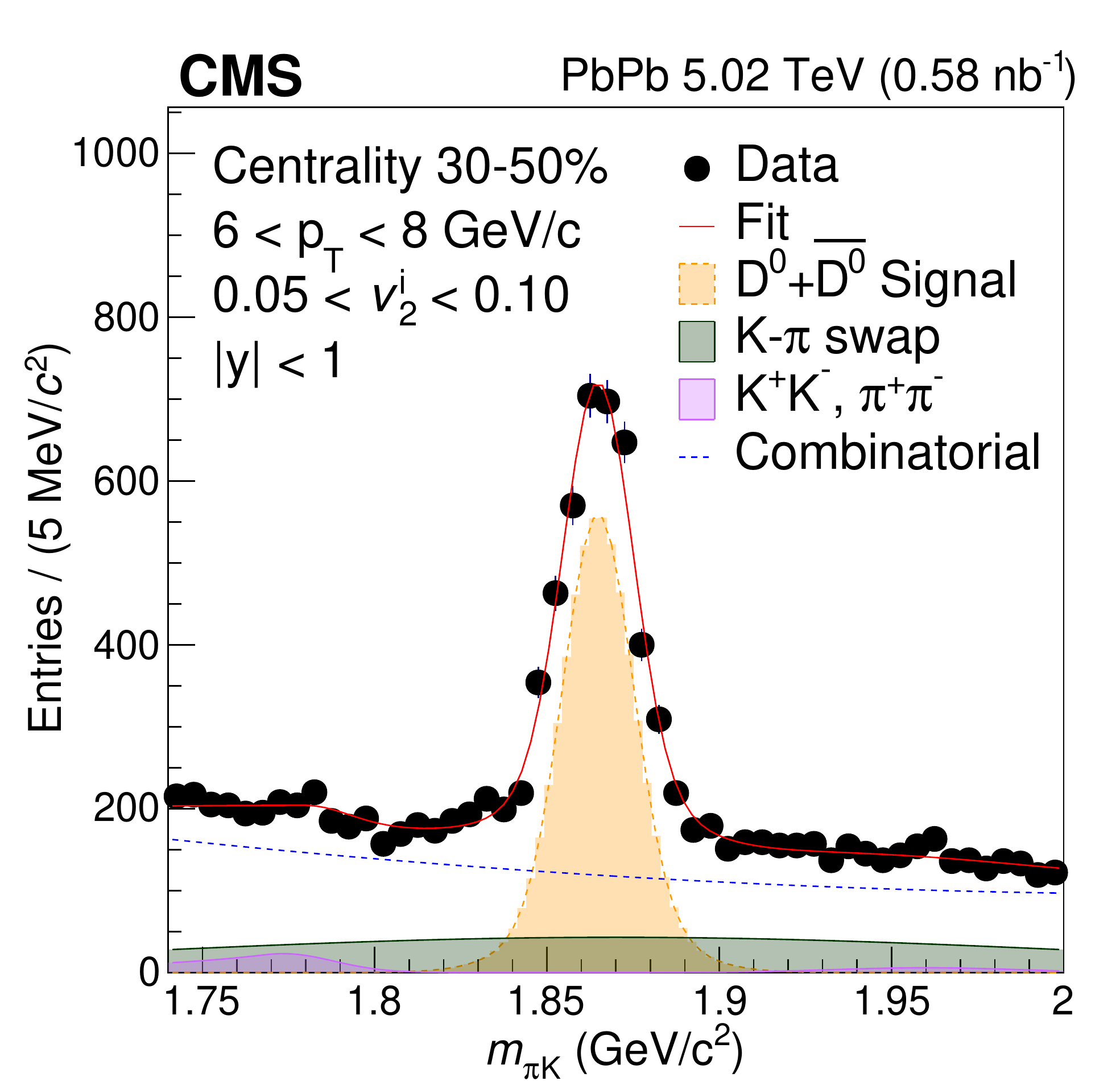}
\includegraphics[width=0.45\textwidth]{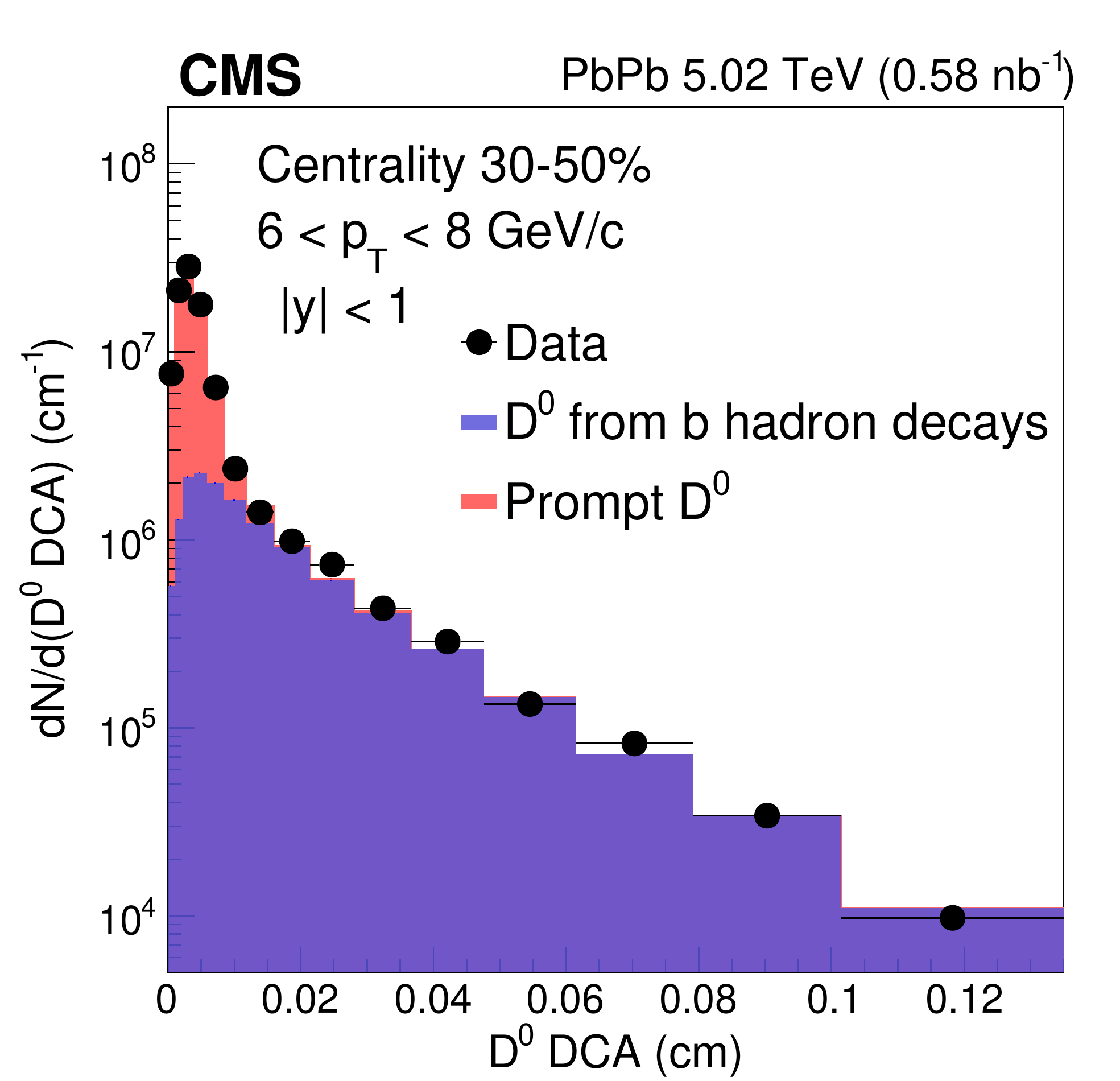}
\caption{An example of the fit to the invariant mass spectrum (left panel) and an example of the template fit of the inclusive \PDz meson yields, extracted as a function of \DCA (right panel). The former fit is used for determining the total \PDz yields and the latter for determining the fraction of nonprompt \PDz mesons.}
\label{fig:massfit}
\end{figure*}

The $\minv$ distributions are fitted with five components, as shown in Fig.~\ref{fig:massfit}: 
the sum of two Gaussian functions with the same mean but different
widths for the \PDz signal, $S(\minv)$; an additional Gaussian function to describe the
invariant mass shape of those \PDz candidates that were given an incorrect mass assignment when swapping 
the pion and kaon designations, $SW(\minv)$; two Crystal Ball functions~\cite{CrystalBallRef} to describe
the processes $\Dtopipi$ ($S(m_{\PGpp\PGpm})$) and $\DtoKK$ ($S(m_{\PKp\PKm})$);
and a third-order polynomial
to model the combinatorial background, $Bkg(\minv)$.
The contributions from the processes $\Dtopipi$ and $\DtoKK$
are also resulting from using an incorrect \PK or \PGp assignment. 
The ratio between the $S(\minv)$ and $SW(\minv)$ function yields is fixed according to values obtained from simulated events.
The widths and means of the $S(\minv)$, $SW(\minv)$, and Crystal Ball functions are initialized using results obtained from simulation studies.
To account for potential differences in momentum resolution between the model and the data, the widths are allowed to vary during the data fitting process.
Specifically, a single scale factor, common to all three functions, is employed to adjust the component of the widths arising from momentum resolution. 
In addition, the means are also allowed to be different between the simulation and data.

The fit of the invariant mass spectrum gives the yield for inclusive \PDz mesons.
For the extraction 
of the two individual components (prompt and nonprompt), a similar procedure as in Refs.~\cite{Sirunyan:2021bvns,Sirunyan:2019braa} is followed.
Distributions of the \DCA 
between the \PDz meson trajectory and the primary vertex
are fitted with a linear combination of prompt and nonprompt \PDz \DCA templates obtained from MC in each bin of \pt, centrality, and $\vn^i$. 
The data distribution was obtained by extracting the \PDz signal from fits to the invariant mass distribution in each \DCA interval.
In generating the simulated templates, the difference of the \DCA values with and without detector effects for each \PDz meson is scaled. 
The same scale factor is chosen for both prompt and nonprompt \PDz mesons, with its value found by minimizing the $\chi^2$ of the combined template fit.

The left panel of Fig.~\ref{fig:massfit} shows an example of a fit to the mass spectrum for \PDz candidates in the \pt interval 6--8\GeVc for the centrality class 30--50\% and the SP range $0.05<v_2^i<0.10$. 
The right panel shows an example of the template fit of the inclusive \PDz meson yields, extracted as a function of \DCA in the same \PDz kinematic region, but averaged over all \vn.

\section{Systematic uncertainties}
\label{sec:syst}

Systematic uncertainties are estimated by varying the conditions and procedures of the \vn measurements.
The absolute differences between the nominal and the alternative analyses are considered as one root-mean-square deviations.
In the fit to \minv, the functional form of the combinatorial background was varied using 
a second-order polynomial and an exponential function.
In the 0--10\% centrality bin, the uncertainty decreases with \pt up to 5\GeVc, remains constant from 5 to 10\GeVc, and increases above 10\GeVc. 
In more peripheral collisions, the uncertainty from this source increases monotonically with \pt over the entire \pt range. 
The mass shape for the signal region is varied by using a triple-Gaussian function and negligible differences are observed. 
The effect of the efficiency is studied by
changing the efficiency correction values. In this alternative approach, the efficiency is based solely on simulation 
and does not consider any potential differences between the MC and data spectra.
The largest effect is observed at low \pt. 
The systematic uncertainty from the determination of the nonprompt \PDz meson fraction is evaluated by performing the template fit using an alternative variable, the \DCA significance (\DCA/(\DCA uncertainty)). 
The differences observed between data and simulation for the \DCA measurement and uncertainty are correlated, as they come from the momentum resolution of the same particles. 
Therefore, this effect is partially canceled in the \DCA significance, obtained by dividing the \DCA by its uncertainty.
However, as \DCA significance arises from the \PDz reconstruction, 
the widths of the distributions cannot be scaled to further match the data, in the same way it was done for \DCA.
The difference in results between the \DCA and \DCA significance template fits is quoted as a systematic uncertainty.
This variation of the template fit gives the largest contribution to the systematic uncertainty, especially at the highest \pt in central collisions.
The effect of the BDT selection is estimated from the MC samples by comparing the results with and without applying it, 
in bins of $\vn^i$, and taking the difference as a systematic uncertainty.
No \pt dependence is observed for this source.
Table~\ref{tab:summary_sys_v2_v3} summarizes 
the estimates of the systematic uncertainties for $\vtwo$ and $\vthree$ from each source.
The total systematic uncertainty is obtained by adding individual uncertainties in quadrature. 

\begin{table*}[h!t]
 \topcaption{Summary of systematic uncertainties in the absolute differences between the nominal and the alternative analyses for $\vtwo$ and $\vthree$.}
 \label{tab:summary_sys_v2_v3}
 \centering 
 \begin{tabular}{lllllll}
  \hline
 Uncertainty source                       & \vtwo  & \vthree    \\
  \hline
  Bkg. mass PDF                           & 0.002--0.012 & 0.001--0.005 \\
  \PDz efficiency correction             &  0.0001--0.0005              & 0.0001--0.0002  \\
  \DCA template fit                           & 0.006--0.013   & 0.002--0.008    \\
  BDT selection                           & 0.001   & 0.006    \\
  \hline
 \end{tabular}
\end{table*}

\section{Results}
\label{sec:res}

\begin{figure*} [ht]
    \centering
    \includegraphics[width=0.98\textwidth]{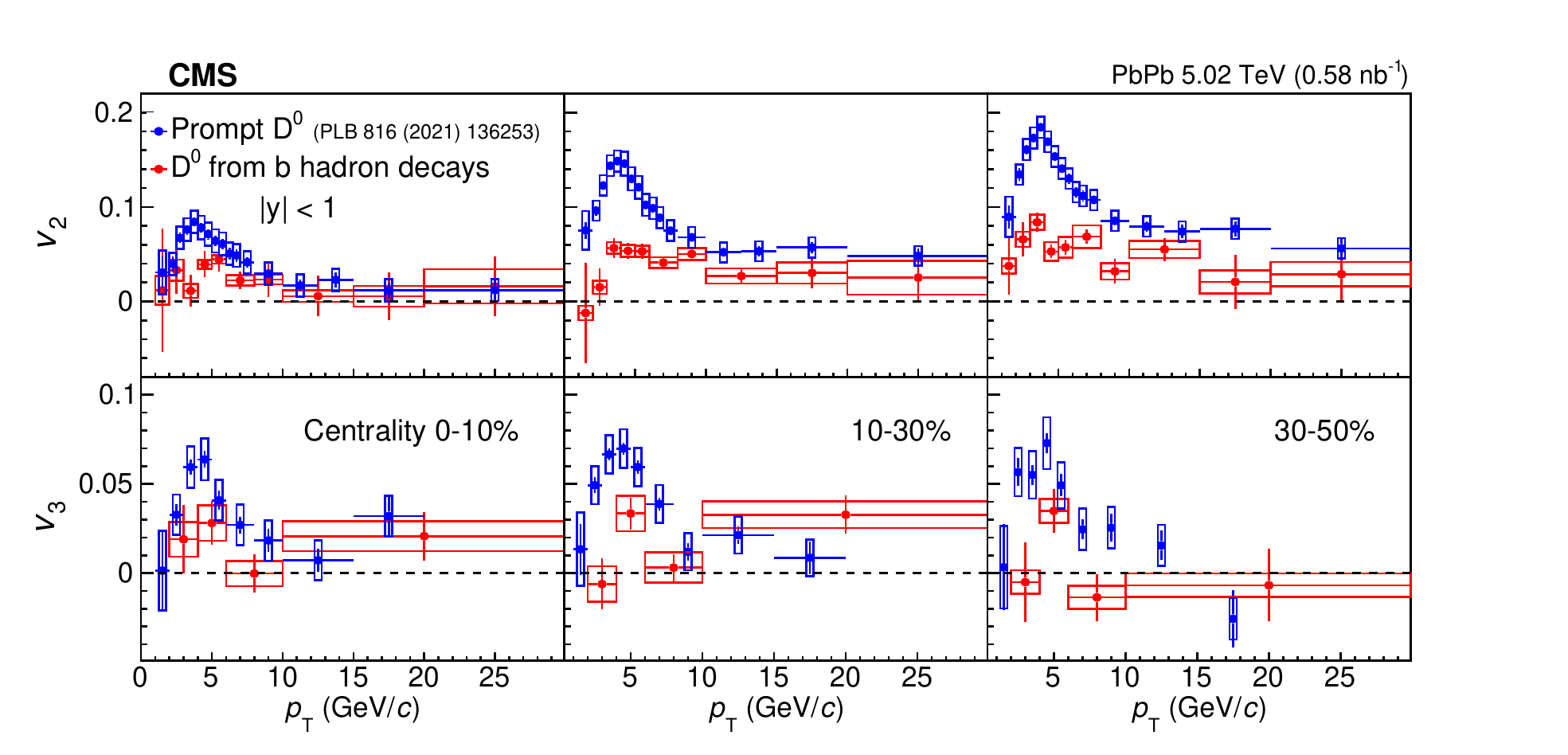}
    \caption{The elliptic,  \vtwo (upper panels), and the triangular, \vthree (lower panels), flow coefficients of nonprompt and prompt (from Ref.~\cite{Sirunyan:2020dvn}) \PDz mesons as functions of their \pt and in three bins of centrality. The bars and the boxes represent statistical and systematic uncertainties, respectively.}
    \label{fig:vn_prompt}
\end{figure*}

The nonprompt \PDz meson flow harmonics are presented in Fig.~\ref{fig:vn_prompt} together with values for prompt \PDz mesons from Ref.~\cite{Sirunyan:2020dvn}.
The results, averaged in the range $2 < \pt < 10\GeVc$, of $\vtwo = 0.0289 \pm 0.0069$, $0.0477 \pm 0.0054$, and $0.060 \pm 0.010$ in the centrality intervals 0--10\%, 10--30\%, and 30--50\%, respectively, 
indicate nonzero values of elliptic flow for B mesons. 
The \vtwo coefficient in the case of \PQb hadron daughters has its maximum value at \pt of ${\sim}5\GeVc$ 
and is significantly lower than in the prompt \PDz meson case. 
This difference becomes more pronounced going from central to peripheral collisions.
These results for the mass ordering of collective flow extend the previously observed relationship between charm and light flavor hadrons, where the flow of charm hadrons has a lower magnitude than that for light flavor particles.
Measurements also suggest an increase of \vtwo towards peripheral collisions, as in the case of light hadrons. 
This observation agrees with the paradigm of flow as a consequence of initial space anisotropy~\cite{Alver:2010prc}.
The elliptic flow of nonprompt \PDz mesons is found to be
of the same magnitude as what the ATLAS Collaboration observed for nonprompt muons~\cite{Aad:2020bvn}.

The nonprompt \vthree coefficient results have large statistical uncertainties so that neither the \pt nor the centrality dependence can be determined. 
As was the case for \vtwo, the triangular flow signal is weaker for nonprompt than for prompt \PDz mesons, but the difference is not as large as that for the elliptic flow.
Measurements, averaged over all three centrality intervals, suggest positive \vthree for nonprompt \PDz mesons in the $4<\pt<6\GeVc$ range with a significance of 3.0 standard deviations. 
The ATLAS Collaboration \vthree results for \PQb hadrons decaying into muons with $\pt>4\GeVc$ were found to be consistent with zero~\cite{Aad:2020bvn}.

\begin{figure*} [ht]
   \centering
   \includegraphics[width=0.98\textwidth]{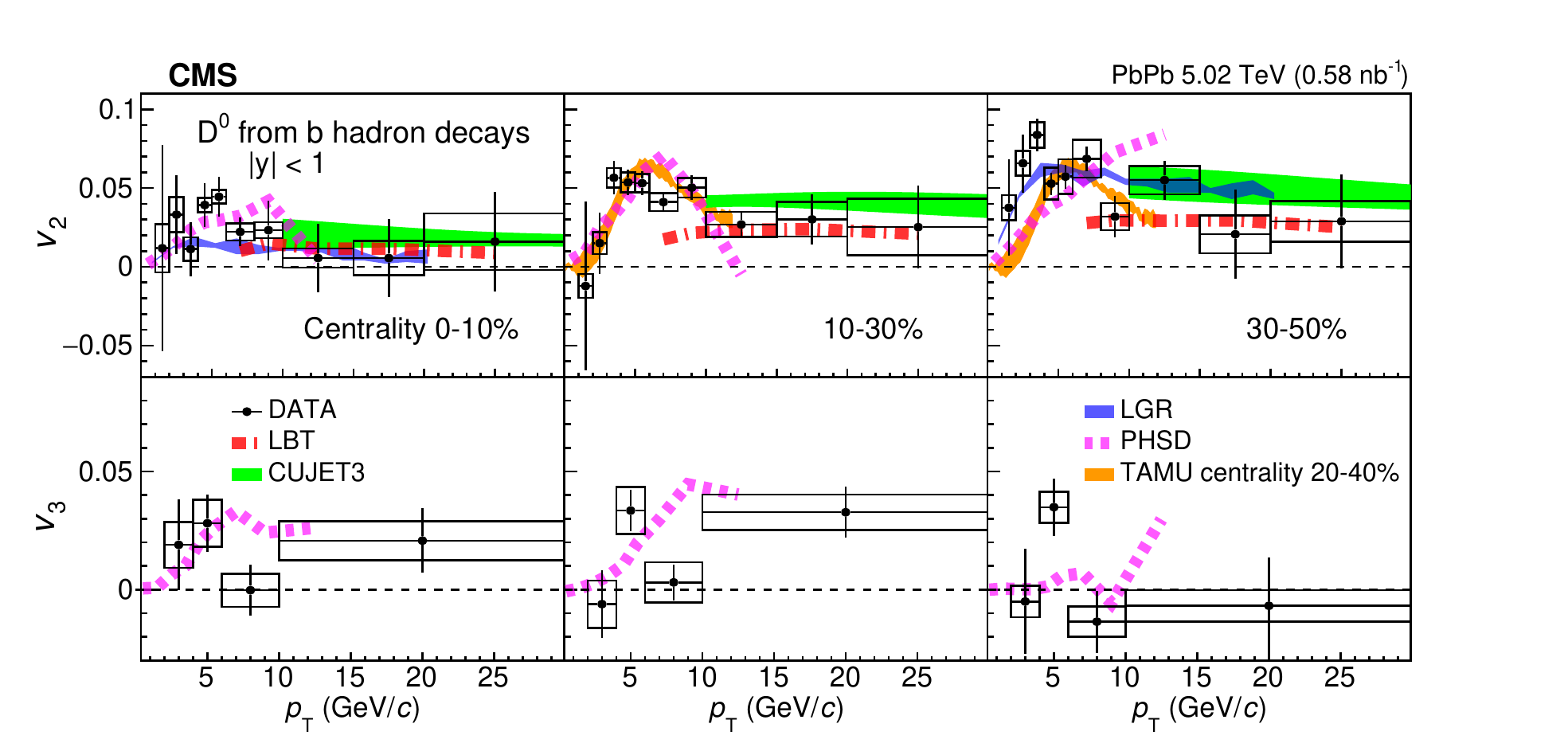}
   \caption{The elliptic, \vtwo (upper panel), and the triangular, \vthree (lower panel), flow coefficients of nonprompt \PDz mesons as functions of their \pt and in three bins of centrality. The bars and the boxes represent statistical and systematic uncertainties, respectively. The colored bands show theoretical predictions~\cite{Song:2015prc, Cao:2016prc, Xing:2020plb, He:2014tamu, Li:2019lgr, Li:2020lgr, Shi:2018cuj, Shi:2019cuj}.}
   \label{fig:v2v3_data}
\end{figure*}

Figure~\ref{fig:v2v3_data} shows the measured nonprompt \PDz meson \vn coefficients compared with theoretical calculations that have different modeling of the \PQb quark flow. 
The PHSD model is a microscopic off-shell transport model based on a Boltzmann transport equation approach, and includes only collisional energy loss~\cite{Song:2015prc}. 
The TAMU model computes the space-time evolution of the heavy-quark phase space distribution in the QGP using the Fokker--Planck equation, implemented via Langevin dynamics, and also has no radiative energy loss~\cite{He:2014tamu}.
The LGR model uses the Langevin approach with gluon radiation, emphasizing the in-medium diffusion dynamics. 
This model aims to describe heavy-quark evolutions from low to intermediate \pt range, $\pt<15\GeVc$~\cite{Li:2019lgr, Li:2020lgr}. 
The LBT model is based on a linearized Boltzmann approach coupled to a hydrodynamic background, and employs both collisional and radiative energy loss~\cite{Cao:2016prc, Xing:2020plb}. 
The CUJET3.0 is essentially a jet energy loss framework based on a nonperturbative QGP medium~\cite{Shi:2018cuj, Shi:2019cuj}. 
The LBT and CUJET3.0 approaches are applicable only for the high-\pt range and do not have predictions for lower momenta. 
The CUJET3.0 model implements hadronization through a parton fragmentation function, while all other models combine coalescence and fragmentation.
All models can qualitatively describe the \pt dependence of the data. The LGR model reproduces the measurements for the centrality 30--50\% and in the low- and the intermediate-\pt ranges, 
where \PQb quarks are expected to follow Brownian motion, i.e., small random momentum fluctuations caused by collisions with thermal particles. 
For the central events, 0--10\%, in the range $4 < \pt <10 \GeVc$, the results from this model lie systematically below the data. 
In the same \pt range, the PHSD model provides a good description of the data, for centralities 0--10\% and 10--30\%, but it cannot describe the data at higher centralities. 
In contrast, this same PHSD model underpredicts the \vtwo Fourier coefficients for \PDz mesons that are not the product of \PQb hadron decays~\cite{Sirunyan:2020dvn}. 
The TAMU model has predictions in the centrality range 20--40\%, 
therefore only an indirect comparison with data is possible, but the predictions describe well the measurement performed in the centrality range 10--30\%. 
At higher \pt, where the anisotropy is driven by the path-length dependence of parton energy loss, 
the LBT model gives lower predictions than CUJET3.0, but both models match data within uncertainties. 
The PHSD, LBT, and LGR models account for event-by-event fluctuations of the initial geometry, but PHSD is the only model that has predictions for \vthree coefficients. 
While precision is limited for both the model and the data, they both reach similar maximum values in all centralities.  

\section{Summary}
\label{sec:sum}

In summary, the elliptic (\vtwo) and triangular (\vthree) flow harmonics of \PDz mesons that originate in \PQb hadron decays (nonprompt \PDz mesons) 
are measured in lead-lead collisions at $\sqrtsNN=5.02\TeV$. 
The \vtwo results show a weak transverse momentum (\pt) dependence and suggest a slight increase for more peripheral collisions. 
An indication of a nonzero \vthree coefficient is found for nonprompt \PDz mesons with $4<\pt<6\GeVc$. 
The magnitudes of the flow coefficients are lower for nonprompt \PDz 
than for prompt \PDz mesons. 
This magnitude difference is more pronounced in the case of \vtwo. 
The qualitative agreement between the results and theoretical models, which include a mass hierarchy in quark interactions with the quark-gluon plasma, extends our understanding of heavy quark interactions with the medium.

\begin{acknowledgments}
We congratulate our colleagues in the CERN accelerator departments for the excellent performance of the LHC and thank the technical and administrative staffs at CERN and at other CMS institutes for their contributions to the success of the CMS effort. In addition, we gratefully acknowledge the computing centers and personnel of the Worldwide LHC Computing Grid and other centers for delivering so effectively the computing infrastructure essential to our analyses. Finally, we acknowledge the enduring support for the construction and operation of the LHC, the CMS detector, and the supporting computing infrastructure provided by the following funding agencies: BMBWF and FWF (Austria); FNRS and FWO (Belgium); CNPq, CAPES, FAPERJ, FAPERGS, and FAPESP (Brazil); MES and BNSF (Bulgaria); CERN; CAS, MoST, and NSFC (China); MINCIENCIAS (Colombia); MSES and CSF (Croatia); RIF (Cyprus); SENESCYT (Ecuador); MoER, ERC PUT and ERDF (Estonia); Academy of Finland, MEC, and HIP (Finland); CEA and CNRS/IN2P3 (France); BMBF, DFG, and HGF (Germany); GSRI (Greece); NKFIH (Hungary); DAE and DST (India); IPM (Iran); SFI (Ireland); INFN (Italy); MSIP and NRF (Republic of Korea); MES (Latvia); LAS (Lithuania); MOE and UM (Malaysia); BUAP, CINVESTAV, CONACYT, LNS, SEP, and UASLP-FAI (Mexico); MOS (Montenegro); MBIE (New Zealand); PAEC (Pakistan); MES and NSC (Poland); FCT (Portugal); MESTD (Serbia); MCIN/AEI and PCTI (Spain); MOSTR (Sri Lanka); Swiss Funding Agencies (Switzerland); MST (Taipei); MHESI and NSTDA (Thailand); TUBITAK and TENMAK (Turkey); NASU (Ukraine); STFC (United Kingdom); DOE and NSF (USA).
    
\hyphenation{Rachada-pisek} Individuals have received support from the Marie-Curie program and the European Research Council and Horizon 2020 Grant, contract Nos.\ 675440, 724704, 752730, 758316, 765710, 824093, 884104, and COST Action CA16108 (European Union); the Leventis Foundation; the Alfred P.\ Sloan Foundation; the Alexander von Humboldt Foundation; the Belgian Federal Science Policy Office; the Fonds pour la Formation \`a la Recherche dans l'Industrie et dans l'Agriculture (FRIA-Belgium); the Agentschap voor Innovatie door Wetenschap en Technologie (IWT-Belgium); the F.R.S.-FNRS and FWO (Belgium) under the ``Excellence of Science -- EOS" -- be.h project n.\ 30820817; the Beijing Municipal Science \& Technology Commission, No. Z191100007219010; the Ministry of Education, Youth and Sports (MEYS) of the Czech Republic; the Hellenic Foundation for Research and Innovation (HFRI), Project Number 2288 (Greece); the Deutsche Forschungsgemeinschaft (DFG), under Germany's Excellence Strategy -- EXC 2121 ``Quantum Universe" -- 390833306, and under project number 400140256 - GRK2497; the Hungarian Academy of Sciences, the New National Excellence Program - \'UNKP, the NKFIH research grants K 124845, K 124850, K 128713, K 128786, K 129058, K 131991, K 133046, K 138136, K 143460, K 143477, 2020-2.2.1-ED-2021-00181, and TKP2021-NKTA-64 (Hungary); the Council of Science and Industrial Research, India; the Latvian Council of Science; the Ministry of Education and Science, project no. 2022/WK/14, and the National Science Center, contracts Opus 2021/41/B/ST2/01369 and 2021/43/B/ST2/01552 (Poland); the Funda\c{c}\~ao para a Ci\^encia e a Tecnologia, grant CEECIND/01334/2018 (Portugal); the National Priorities Research Program by Qatar National Research Fund; MCIN/AEI/10.13039/501100011033, ERDF ``a way of making Europe", and the Programa Estatal de Fomento de la Investigaci{\'o}n Cient{\'i}fica y T{\'e}cnica de Excelencia Mar\'{\i}a de Maeztu, grant MDM-2017-0765 and Programa Severo Ochoa del Principado de Asturias (Spain); the Chulalongkorn Academic into Its 2nd Century Project Advancement Project, and the National Science, Research and Innovation Fund via the Program Management Unit for Human Resources \& Institutional Development, Research and Innovation, grant B05F650021 (Thailand); the Kavli Foundation; the Nvidia Corporation; the SuperMicro Corporation; the Welch Foundation, contract C-1845; and the Weston Havens Foundation (USA). 
\end{acknowledgments}

\bibliography{auto_generated} 
\cleardoublepage \appendix\section{The CMS Collaboration \label{app:collab}}\begin{sloppypar}\hyphenpenalty=5000\widowpenalty=500\clubpenalty=5000
\cmsinstitute{Yerevan Physics Institute, Yerevan, Armenia}
{\tolerance=6000
A.~Tumasyan\cmsAuthorMark{1}\cmsorcid{0009-0000-0684-6742}
\par}
\cmsinstitute{Institut f\"{u}r Hochenergiephysik, Vienna, Austria}
{\tolerance=6000
W.~Adam\cmsorcid{0000-0001-9099-4341}, J.W.~Andrejkovic, T.~Bergauer\cmsorcid{0000-0002-5786-0293}, S.~Chatterjee\cmsorcid{0000-0003-2660-0349}, K.~Damanakis\cmsorcid{0000-0001-5389-2872}, M.~Dragicevic\cmsorcid{0000-0003-1967-6783}, A.~Escalante~Del~Valle\cmsorcid{0000-0002-9702-6359}, P.S.~Hussain\cmsorcid{0000-0002-4825-5278}, M.~Jeitler\cmsAuthorMark{2}\cmsorcid{0000-0002-5141-9560}, N.~Krammer\cmsorcid{0000-0002-0548-0985}, L.~Lechner\cmsorcid{0000-0002-3065-1141}, D.~Liko\cmsorcid{0000-0002-3380-473X}, I.~Mikulec\cmsorcid{0000-0003-0385-2746}, P.~Paulitsch, F.M.~Pitters, J.~Schieck\cmsAuthorMark{2}\cmsorcid{0000-0002-1058-8093}, R.~Sch\"{o}fbeck\cmsorcid{0000-0002-2332-8784}, D.~Schwarz\cmsorcid{0000-0002-3821-7331}, M.~Sonawane\cmsorcid{0000-0003-0510-7010}, S.~Templ\cmsorcid{0000-0003-3137-5692}, W.~Waltenberger\cmsorcid{0000-0002-6215-7228}, C.-E.~Wulz\cmsAuthorMark{2}\cmsorcid{0000-0001-9226-5812}
% ...
\par}
\vskip\cmsinstskip
\dag:~Deceased\\
$^{1}$Also at Yerevan State University, Yerevan, Armenia\\
$^{2}$Also at TU Wien, Vienna, Austria\\
% ...
\end{sloppypar}
\end{document}